\documentclass[%
superscriptaddress,
preprint,
 amsmath,amssymb,
 aps,
]{revtex4-1}

\usepackage{graphicx}
\usepackage{dcolumn}
\usepackage{multirow}
\usepackage{bm}
\usepackage{xcolor}
\usepackage{qcircuit}
\usepackage{braket}
\usepackage{float}
\usepackage{algorithm}
\usepackage{algorithmicx}
\usepackage{algpseudocode}
\usepackage{mathtools}
\usepackage{hyperref}

\renewcommand{\figureautorefname}{Figure~\negthinspace}

\renewcommand{\tableautorefname}{Table~\negthinspace}
\renewcommand{\sectionautorefname}{Section~\negthinspace}

\begin{document}


\title{Asynchronous training of quantum reinforcement learning}

\author{Samuel Yen-Chi Chen}
\affiliation{%
 Wells Fargo
}%

\date{\today}

\begin{abstract}
The development of quantum machine learning (QML) has received a lot of interest recently thanks to developments in both quantum computing (QC) and machine learning (ML). One of the ML paradigms that can be utilized to address challenging sequential decision-making issues is reinforcement learning (RL). It has been demonstrated that classical RL can successfully complete many difficult tasks.
A leading method of building quantum RL agents relies on the variational quantum circuits (VQC). However, training QRL algorithms with VQCs requires significant amount of computational resources. This issue hurdles the exploration of various QRL applications.
In this paper, we approach this challenge through asynchronous training QRL agents. Specifically, we choose the asynchronous training of advantage actor-critic variational quantum policies.
We demonstrate the results via numerical simulations that within the tasks considered, the asynchronous training of QRL agents can reach performance comparable to or superior than classical agents with similar model sizes and architectures.

\end{abstract}

\maketitle


\section{\label{sec:Indroduction}Introduction}

Quantum computing (QC) has been posited as a means of achieving computational superiority for certain tasks that classical computers struggle to solve \cite{nielsen2010quantum}. Despite this potential, the lack of error-correction in current quantum computers has made it challenging to effectively implement complex quantum circuits on these "noisy intermediate-scale quantum" (NISQ) devices \cite{preskill2018quantum}. To harness the quantum advantages offered by NISQ devices, the development of specialized quantum circuit architectures is necessary.

Recent advances in the hybrid quantum-classical computing framework \cite{bharti2022noisy} that utilizes both classical and quantum computing. Under this approach, certain computational tasks that are expected to benefit from quantum processing are executed on a quantum computer, while others, such as gradient calculations, are performed on classical computers. This hybrid approach aims to take advantage of the strengths of both types of computing to address a wide range of tasks.
Hybrid algorithms that utilize variational quantum circuits (VQC) have proven to be effective in a variety of machine learning tasks. VQCs are a subclass of quantum circuits that possess tunable parameters, and their incorporation into QML models has demonstrated success in a wide range of tasks \cite{bharti2022noisy,cerezo2021variational}.

Reinforcement learning (RL) is a branch of machine learning that deals with sequential decision making tasks. Deep neural network-based RL has achieved remarkable results in complicated tasks with human-level \cite{mnih2015human} or super-human performance \cite{silver2017mastering}. However, quantum RL is a developing field with many unresolved issues and challenges.
The majority of existing quantum RL models are based on VQC \cite{chen19,lockwood2020reinforcement,skolik2021quantum,jerbi2021variational,hsiao2022unentangled}. Although these models have been shown to perform well in a variety of benchmark tasks, training them requires a significant amount of computational resources. The long training time limits the exploration of quantum RL's broad application possibilities.
We propose an asynchronous training framework for quantum RL agents in this paper. We focus on the asynchronous training of advantage actor-critic quantum policies using multiple instances of agents running in parallel.

We show, using numerical simulations, that quantum models may outperform or be similar to classical models in the various benchmark tasks considered. Furthermore, the suggested training approach has the practical advantage of requiring significantly less time for training, allowing for more quantum RL applications.
%
%

The structure of this paper is as follows: In \sectionautorefname{\ref{sec:RelevantWorks}}, we provide an overview of relevant prior work and compare our proposal to these approaches. In \sectionautorefname{\ref{sec:RL}}, we provide a brief overview of the necessary background in reinforcement learning. In \sectionautorefname{\ref{sec:VQC}}, we introduce the concept of variational quantum circuits (VQCs), which serve as the building blocks of our quantum reinforcement learning agents. In \sectionautorefname{\ref{sec:QuantumA3C}}, we present our proposed quantum A3C framework. In \sectionautorefname{\ref{sec:ExpAndResults}}, we describe our experimental setup and present our results. Finally, in \sectionautorefname{\ref{sec:Conclusion}}, we offer some concluding remarks.
\section{\label{sec:RelevantWorks}Relevant Works}
The work that gave rise to quantum reinforcement learning (QRL) \cite{meyer2022survey} may be traced back to \cite{dong2008quantum}. However, the framework demands a quantum environment, which may not be met in most real-world situations. Further studies utilizing Grover-like methods include \cite{wiedemann2022quantum,sannia2022hybrid}. Quantum linear system solvers are also used to implement quantum policy iteration \cite{cherrat2022quantum}. We will concentrate on recent advancements in VQC-based QRL dealing with classical environments.
The first VQC-based QRL \cite{chen19}, which is the quantum version of deep $Q$-learning (DQN), considers discrete observation and action spaces in the testing environments such as Frozen-Lake and Cognitive-Radio. Later, more sophisticated efforts in the area of quantum DQN take into account continuous observation spaces like Cart-Pole \cite{lockwood2020reinforcement,skolik2021quantum}. A further development along this direction includes the using of quantum recurrent neural networks such as QLSTM as the value function approximator \cite{chen2022quantum} to tackle challenges such as partial observability or environments requiring longer memory of previous steps.
Various methods such as hybrid quantum-classical linear solver are developed to find value functions \cite{Chih-ChiehCHEN2020}. A further improvement of DQN which can improve the agent convergence such as Double DQN (DDQN) are also implemented within VQC framework in the work \cite{heimann2022quantum}, in which the authors apply QRL to solve robot navigation task. 
Recent advances in QRL have led to the development of frameworks that aim to learn policy functions, denoted as $\pi$, directly. These frameworks are able to learn the optimal policy for a given problem, in addition to learning value functions such as the $Q$-function. For example, the paper \cite{jerbi2021variational} describes the quantum policy gradient RL through the use of REINFORCE algorithm. Then, the work \cite{hsiao2022unentangled} consider an improved policy gradient algorithm called PPO with VQCs and show that even with a small number of parameters, quantum models can outperform their classical counterparts. Provable quantum advantages of policy gradient are shown in the work \cite{jerbi2022quantum}. Additional research, such as the work in \cite{meyer2022quantum}, has explored the impact of various post-processing methods for VQC on the performance of quantum policy gradients.
Several improved quantum policy gradient algorithms have been proposed in recent years, including actor-critic \cite{schenk2022hybrid} and soft actor-critic (SAC) \cite{lan2021variational,acuto2022variational}. These modifications seek to further improve the efficiency and effectiveness of QRL methods.
QRL has also been applied to the field of quantum control \cite{sequeira2022variational} and has been extended to the multi-agent setting \cite{yun2022quantum,yan2022multiagent,yun2022quantum2}.
The work \cite{chen2022variational} were the first to explore the use of evolutionary optimization for QRL. In their work, multiple agents were initialized and run in parallel, with the top performing agents being selected as parents to generate the next generation of agents.
In the work \cite{wu2020quantum}, the authors studied the use of advanced quantum policy gradient methods, such as the deep deterministic policy gradient (DDPG) algorithm, for QRL in continuous action spaces. 

In this work, we extend upon previous research on quantum policy gradient \cite{jerbi2021variational,hsiao2022unentangled,schenk2022hybrid} by introducing an asynchronous training method for quantum policy learning. While previous approaches have employed single-threaded training, our method utilizes an asynchronous approach, which may offer practical benefits such as reduced training time through the use of multi-core CPU computing resources and the potential for utilizing multiple quantum processing units (QPUs) in the future. Our approach shares some similarities with the evolutionary QRL method presented in \cite{chen2022variational}, which also utilizes parallel computing resources. However, our approach differs in that individual agents can share their gradients directly with the shared global gradient asynchronously, rather than waiting for all agents to finish before calculating fitness and creating the next generation of agents. This characteristic may further improve the efficiency of the training process. These contributions represent a novel advancement in the field of quantum reinforcement learning.

\section{\label{sec:RL}Reinforcement Learning}
\emph{Reinforcement learning} (RL) is a machine learning framework in which an \emph{agent} learns to accomplish a given goal by interacting with an \emph{environment} $\mathcal{E}$ in discrete time steps~\cite{sutton2018reinforcement}.
The agent observes a \emph{state} $s_t$ at each time step $t$ and then chooses an \emph{action} $a_t$ from the action space $\mathcal{A}$ based on its current \emph{policy} $\pi$. The policy is a mapping from a specific state $s_t$ to the probabilities of choosing one of the actions in $\mathcal{A}$.
After performing the action $a_t$, the agent gets a scalar \emph{reward} $r_t$ and the state of the following time step $s_{t+1}$ from the environment. For episodic tasks, the procedure is repeated across a number of time steps until the agent reaches the terminal state or the maximum number of steps permitted.
Seeing the state $s_t$ along the training process, the agent aims to maximize the expected return, which can be expressed as the value function at state $s$ under policy $\pi$, $V^\pi(s) = \mathbb{E}\left[R_t|s_t = s\right]$, where $R_t = \sum_{t'=t}^{T} \gamma^{t'-t} r_{t'}$ is the \emph{return}, the total discounted reward from time step $t$. The value function can be further expressed as $V^\pi(s) = \sum_{a\in\mathcal{A}} Q^\pi (s,a) \pi(a|s)$, where the \emph{action-value function} or \emph{Q-value function} $ Q^\pi (s,a) = \mathbb{E}[R_t|s_t = s, a]$ is the expected return of choosing an action $a \in \mathcal{A}$ in state $s$ according to the policy $\pi$. The $Q$-learning is RL algorithm to optimize the $Q^\pi (s,a)$ via the following formula
\begin{align}
  Q\left(s_{t}, a_{t}\right) \leftarrow & \, Q\left(s_{t}, a_{t}\right)\nonumber\\
  &+\alpha\left[r_{t}+\gamma \max _{a} Q\left(s_{t+1}, a\right)-Q\left(s_{t}, a_{t}\right)\right].
\end{align}

In contrast to \emph{value-based} reinforcement learning techniques, such as $Q$-learning, which rely on learning a value function and using it to guide decision-making at each time step, \emph{policy gradient} methods focus on directly optimizing a policy function, denoted as $\pi(a|s;\theta)$, parametrized by $\theta$. The parameters $\theta$ are updated through a gradient ascent procedure on the expected total return, $\mathbb{E}[R_{t}]$. A notable example of a policy gradient algorithm is the REINFORCE algorithm, introduced in~\cite{williams1992simple}.
In the standard REINFORCE algorithm, the parameters $\theta$ are updated along the direction $\nabla_{\theta} \log \pi\left(a_{t} | s_{t} ; \theta\right) R_{t}$, which is an unbiased estimate of $\nabla_{\theta} \mathbb{E}\left[R_{t}\right]$. However, this policy gradient estimate often suffers from high variance, making training difficult. To reduce the variance of this estimate while maintaining its unbiasedness, a term known as the \emph{baseline} can be subtracted from the return. This baseline, denoted as $b_{t}(s_{t})$, is a learned function of the state $s_{t}$. The resulting update becomes $\nabla_{\theta} \log \pi\left(a_{t} | s_{t} ; \theta\right)\left(R_{t}-b_{t}\left(s_{t}\right)\right)$.
A common choice for the baseline $b_t(s_t)$ in RL is an estimate of the value function $V^\pi(s_t)$. Using this choice for the baseline often results in a lower variance estimate of the policy gradient \cite{sutton2018reinforcement}. The quantity $R_t - b_t = Q(s_t, a_t) - V(s_t)$ can be interpreted as the \emph{advantage} $A(s_t, a_t)$ of action $a_t$ at state $s_t$. Intuitively, the advantage can be thought of as the "goodness or badness" of action $a_t$ relative to the average value at state $s_t$. 
This approach is known as the advantage actor-critic (A2C) method, where the policy $\pi$ is the actor and the baseline, which is the value function $V$, is the critic~\cite{sutton2018reinforcement}.

The asynchronous advantage actor-critic (A3C) algorithm \cite{mnih2016asynchronous} is a variant of the A2C method that employs multiple concurrent actors to learn the policy through parallelization. Asynchronous training of RL agents involves executing multiple agents on multiple instances of the environment, allowing the agents to encounter diverse states at any given time step. This diminished correlation between states or observations enhances the numerical stability of on-policy RL algorithms such as actor-critic \cite{mnih2016asynchronous}. Furthermore, asynchronous training does not require the maintenance of a large replay memory, thus reducing memory requirements \cite{mnih2016asynchronous}. By harnessing the advantages and gradients computed by a pool of actors, A3C exhibits impressive sample efficiency and robust learning performance, making it a prevalent choice in the realm of reinforcement learning.
\section{\label{sec:VQC}Variational Quantum Circuit}
\begin{figure}[htbp]
\includegraphics[width=0.6\linewidth]{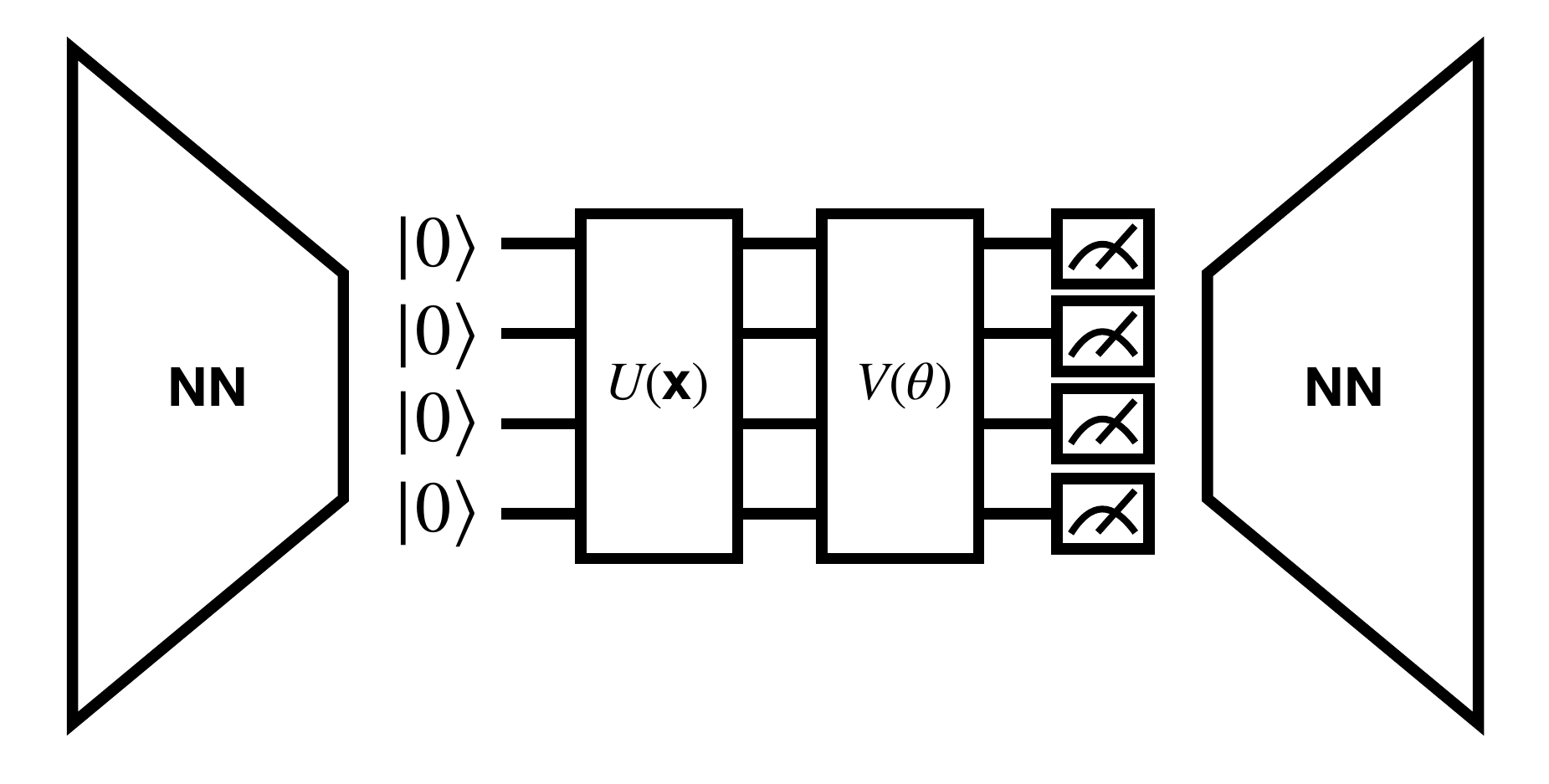}
\caption{{\bfseries Hybrid variational quantum circuit (VQC) architecture.} The hybrid VQC architecture includes a VQC and classical neural networks (NN) before and after the VQC. NN can be used to reduce the dimensionality of the input data and refine the outputs from the VQC.}
\label{fig:hybrid_vqc}
\end{figure}

%
%
Variational quantum circuits (VQCs), also referred to as parameterized quantum circuits (PQCs), are a class of quantum circuits that contain tunable parameters. These parameters can be optimized using various techniques from classical machine learning, including gradient-based and non-gradient-based methods. A generic illustration of a VQC is in the central part of \figureautorefname{\ref{fig:hybrid_vqc}}.

The three primary components of a VQC are the encoding circuit, the variational circuit, and the quantum measurement layer. The encoding circuit, denoted as $U(\textbf{x})$, transforms classical values into a quantum state, while the variational circuit, denoted as $V(\theta)$, serves as the learnable part of the VQC. The quantum measurement layer, on the other hand, is utilized to extract information from the circuit. It is a common practice to repeatedly execute the circuit, also known as "shots," in order to obtain the expectation values of each qubit. A common choice is to use the Pauli-$Z$ expectation values. Instead of being binary integers, the values are received as floats. Additionally, other components, such as additional VQCs or classical components such as DNN, can process the values obtained from the circuit.

The VQC can operate with other classical components such as tensor networks (TN) \cite{chen2022variational,chen2021end,qi2021qtn} and deep neural networks (NN) to perform data pre-processing such as dimensional reduction or post-processing such as scaling. We call such VQCs as \emph{dressed} VQC, as shown in \figureautorefname{\ref{fig:hybrid_vqc}}.
The whole model can be trained in an end-to-end manner via gradient-based \cite{chen2021end,qi2021qtn} or gradient-free methods \cite{chen2022variational}. For the gradient-based methods, the whole model can be represented as a directed acyclic graph (DAG) and then back-propagation can be applied. The success of such end-to-end optimization relies on the capabilities of calculating the quantum gradients such as \emph{parameter-shift} rule \cite{mitarai2018quantum}.
VQC-based QML models have shown success in areas such as classification \cite{mitarai2018quantum,qi2021qtn,chehimi2022quantum,chen2021federated,chen2021end}, natural language processing \cite{yang2020decentralizing,yang2022bert,di2022dawn} and sequence modeling \cite{chen2020quantum,chen2022reservoir}.
\section{\label{sec:QuantumA3C}Quantum A3C}
\begin{figure}[htbp]
\includegraphics[width=0.7\linewidth]{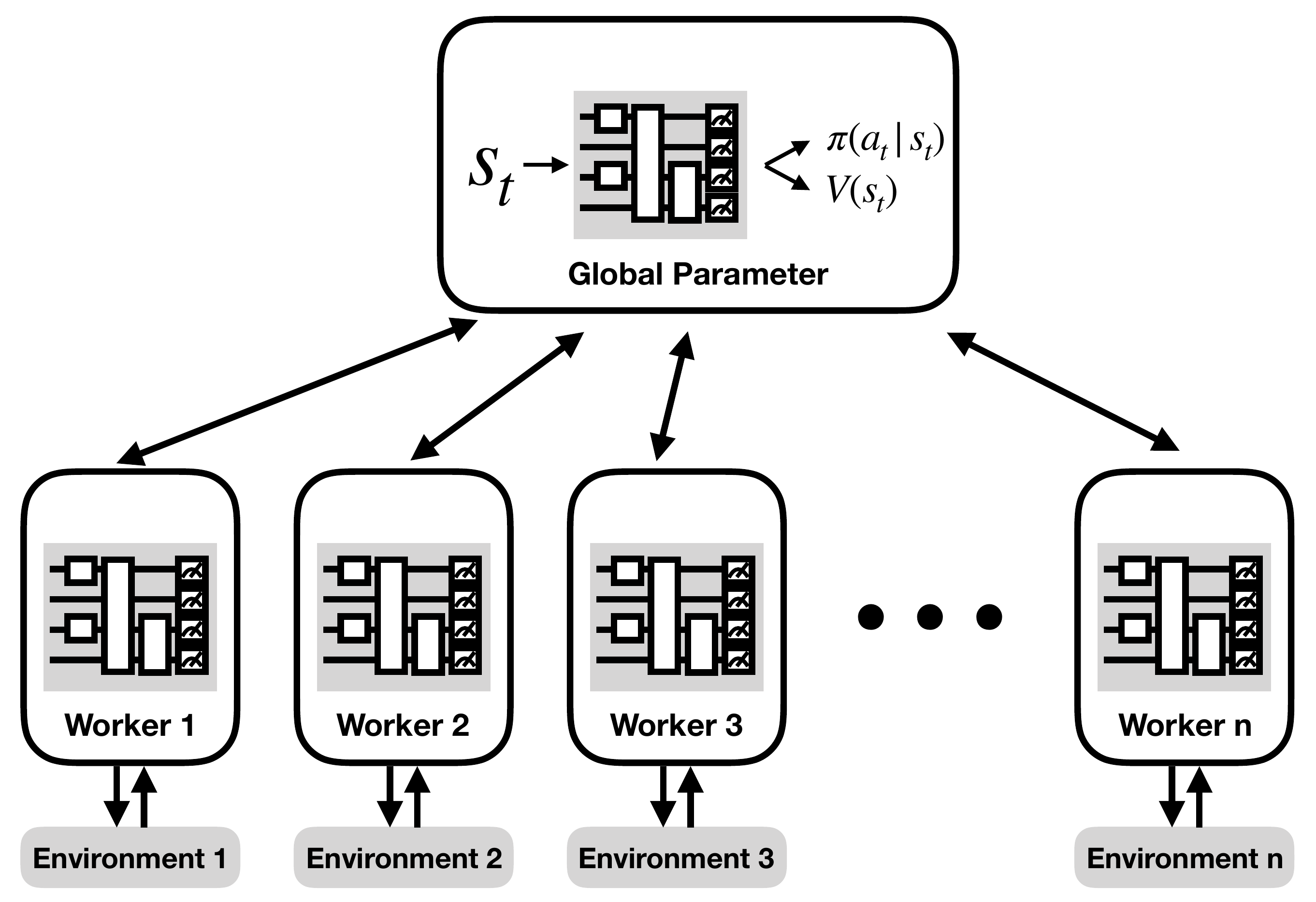}
\caption{{\bfseries Quantum asynchronous advantage actor-critic (A3C) learner.} The proposed quantum A3C includes a global shared parameters and multiple parallel workers. The action generation process within each local agent is performed using the dressed VQC policy and value functions stored in the process-specific memories. Upon meeting certain criteria, the gradients of the local model parameters are uploaded to the global shared memory, where the global model parameters are updated. The updated global model parameters are then immediately downloaded to the local agent that just uploaded its own gradients.}
\label{fig:quantum_a3c}
\end{figure}

The proposed quantum asynchronous advantage actor-critic (QA3C) framework consists of two main components: a \emph{global shared memory} and \emph{process-specific memories} for each agent. The global shared memory maintains the dressed VQC policy and value parameters, which are modified when an individual process uploads its own gradients for parameter updates. Each agent has its own process-specific memory that maintains local dressed VQC policy and value parameters. These local models are used to generate actions during an episode within individual processes. When certain criteria are met, the gradients of the local model parameters are uploaded to the global shared memory, and the global model parameters are modified accordingly. The updated global model parameters are then immediately downloaded to the local agent that just uploaded its own gradients. The overall concept of QA3C is depicted in \figureautorefname{\ref{fig:quantum_a3c}}.

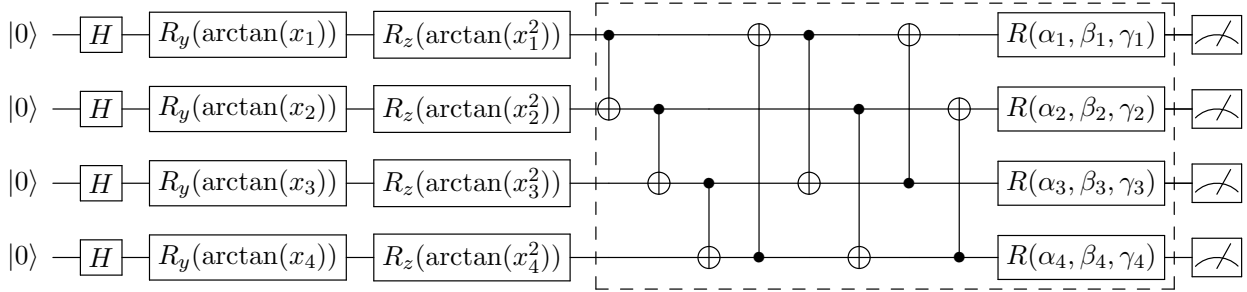
\begin{figure}[htbp]
\begin{center}
\scalebox{0.95}{
\begin{minipage}{10cm}
\Qcircuit @C=1em @R=1em {
\lstick{\ket{0}} & \gate{H} & \gate{R_y(\arctan(x_1))} & \gate{R_z(\arctan(x_1^2))} & \ctrl{1}   & \qw       & \qw      & \targ    & \ctrl{2}   & \qw      & \targ    & \qw      & \gate{R(\alpha_1, \beta_1, \gamma_1)} & \meter \qw \\
\lstick{\ket{0}} & \gate{H} & \gate{R_y(\arctan(x_2))} & \gate{R_z(\arctan(x_2^2))} & \targ      & \ctrl{1}  & \qw      & \qw      & \qw        & \ctrl{2} & \qw      & \targ    & \gate{R(\alpha_2, \beta_2, \gamma_2)} & \meter \qw \\
\lstick{\ket{0}} & \gate{H} & \gate{R_y(\arctan(x_3))} & \gate{R_z(\arctan(x_3^2))} & \qw        & \targ     & \ctrl{1} & \qw      & \targ      & \qw      & \ctrl{-2}& \qw      & \gate{R(\alpha_3, \beta_3, \gamma_3)} & \meter \qw \\
\lstick{\ket{0}} & \gate{H} & \gate{R_y(\arctan(x_4))} & \gate{R_z(\arctan(x_4^2))} & \qw        & \qw       & \targ    & \ctrl{-3}& \qw        & \targ    & \qw      & \ctrl{-2}& \gate{R(\alpha_4, \beta_4, \gamma_4)} & \meter \gategroup{1}{5}{4}{13}{.7em}{--}\qw 
}
\end{minipage}}
\end{center}
\caption{{\bfseries VQC architecture for quantum A3C.} The VQC used here includes $R_{y}$ and $R_{z}$ for encoding classical values $\mathbf{x}$, multiple CNOT gates to entangle qubits, general unitary rotations $R$ and the final measurement. The output of the VQC consists of Pauli-$Z$ expectation values, which are obtained through multiple runs (shots) of the circuit. These values are then processed by classical neural networks for further use. We use a 4-qubit system as an example here, however, it can be enlarge or shrink based on the problem of interest. In this work, the number of qubit is 8.}
\label{fig:basic_vqc_in_qa3c}
\end{figure}

We construct the quantum policy $\pi\left(a_t \mid s_t ; \theta\right)$ and value $V\left(s_t ; \theta_v\right)$ function with the dressed VQC as shown in \figureautorefname{\ref{fig:hybrid_vqc}}, in which the VQC follows the architecture shown in \figureautorefname{\ref{fig:basic_vqc_in_qa3c}}. This VQC architecture has been studied in the work such as quantum recurrent neural networks (QRNN) \cite{chen2020quantum}, quantum recurrent RL \cite{chen2022quantum}, quantum convolutional neural networks \cite{chen2020qcnn}, federated quantum classification \cite{chen2021federated} and has demonstrated superior performance over their classical counterparts under certain conditions. In addition, we employ the classical DNN before and after the VQC to dimensionally reduce the data and fine-tune the outputs from the VQC, respectively. 
The neural network components in this hybrid architecture consist of single-layer networks for dimensionality conversion. Specifically, the network preceding the VQC is a linear layer with an input dimension equal to the size of the observation vector and an output dimension equal to the number of qubits in the VQC. 
The networks following the VQC are linear layers with input dimensions equal to the number of qubits in the VQC and output dimensions equal to the number of actions (for the actor function $\pi\left(a_t \mid s_t ; \theta\right)$) or 1 (for the critic function $V\left(s_t ; \theta_v\right)$). These layers serve to convert the output of the VQC for use in the actor-critic algorithm.
The policy and value function are updated after every $S$ steps or when the agent reaches the terminal state. The details of the algorithm such as the gradient update formulas are presented in Algorithm~\ref{QA3C_alg}.

\section{\label{sec:ExpAndResults}Experiments and Results}
\subsection{Testing Environments}
\subsubsection{Acrobot}

The Acrobot environment from OpenAI Gym \cite{brockman2016openai} 
 consists of a system with two linearly connected links, with one end fixed. The joint connecting the two links can be actuated by applying torques. The goal is to swing the free end of the chain over a predetermined height, starting from a downward hanging position, using as few steps as possible. The observation in this environment is a six-dimensional vector comprising the sine and cosine values of the two rotational joint angles, as well as their angular velocities. The agents are able to take one of three actions: applying $-1$, $0$, or $+1$ torque to the actuated joint. An action resulting in the free end reaching the target height receives a reward of $0$ and terminates the episode. Any action that does not lead to the desired height receives a reward of $-1$. The reward threshold is $-100$.
\begin{figure}[htbp]
\center
\includegraphics[width=0.5\linewidth]{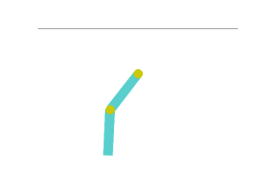}
\caption[Environment: Acrobot]{{\bfseries The Acrobat environment from OpenAI Gym.}
 }
\label{Acrobot_Env}
\end{figure}
\subsubsection{Cart-Pole}
\begin{figure}[htbp]
\center
\includegraphics[width=0.5\linewidth]{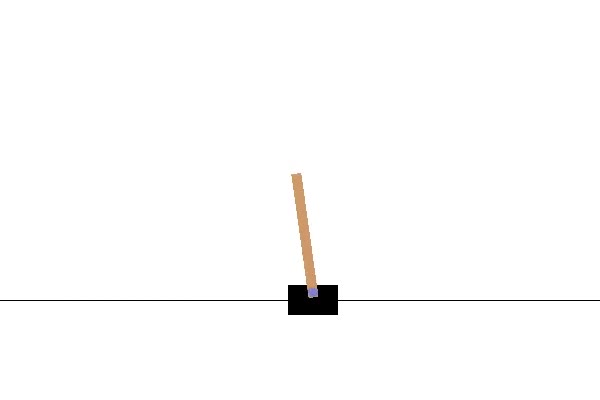}
\caption[Environment: Cart-Pole]{{\bfseries The Cart-Pole environment from OpenAI Gym.}
 }
\label{CartPole_Env}
\end{figure}

Cart-Pole is a commonly used evaluation environment for simple RL models that has been utilized as a standard example with in OpenAI Gym \cite{brockman2016openai} (see~\figureautorefname{\ref{CartPole_Env}}).
A fixed junction connects a pole to a cart traveling horizontally over a frictionless track in this environment. The pendulum initially stands upright, and the aim is to keep it as near to its starting position as possible by moving the cart left and right.
Each time step, the RL agent learns to produce the right action according on the observation it gets.
The observation in this environment is a four dimensional vector $s_t$ containing values of the {cart position, cart velocity, pole angle, and pole velocity at the tip}. 
Every time step where the pole is near to being upright results in a $+1$ award. An episode ends if the pole is inclined more than $15$ degrees from vertical or the cart moves more than $2.4$ units away from the center.
\subsubsection{MiniGrid-SimpleCrossing}
The MiniGrid-SimpleCrossing environment \cite{gym_minigrid} is more sophisticated, with a lot bigger observation input for the RL agent. In this scenario, the RL agent receives a $7 \times 7 \times 3 = 147$ dimensional vector through observation and must choose an action from the action space $\mathcal{A}$, which offers six options. It is important to note that the $147$-dimensional vector is a compact and efficient representation of the environment rather than the real pixels.
There are six actions  {$0$,$\cdots$,$5$} in the action space $\mathcal{A}$ for the agent to choose. They are \textit{turn left}, \textit{turn right}, \textit{move forward}, \textit{pick up an object}, \textit{drop the object being carried} and \textit{toggle}. Only the first three of them are having actual effects in this case. The agent is expected to learn this fact.
In this environment, the agent receives a reward of 1 upon reaching the goal. A penalty is subtracted from this reward based on the formula $1 - 0.9 \times (\textit{number of steps}/\textit{max steps allowed})$, where the maximum number of steps allowed is defined as $4 \times n \times n$, and $n$ is the grid size \cite{gym_minigrid}. In this work, $n$ is set to 9. This reward scheme presents a challenge because it is \emph{sparse}, meaning that the agent does not receive rewards until it reaches the goal.
As shown in \figureautorefname{\ref{SimpleCrossing_Env}}, the agent (shown in red triangle) is expected to find the shortest path from the starting point to the goal (shown in green).
We consider three cases in this environment: MiniGrid-SimpleCrossingS9N1-v0, MiniGrid-SimpleCrossingS9N2-v0 and {MiniGrid-SimpleCrossingS9N3-v0. Here the $N$ represents the number of valid crossings across walls from the starting position to the goal.
\begin{figure}[htbp]
\center
\includegraphics[width=1.\linewidth]{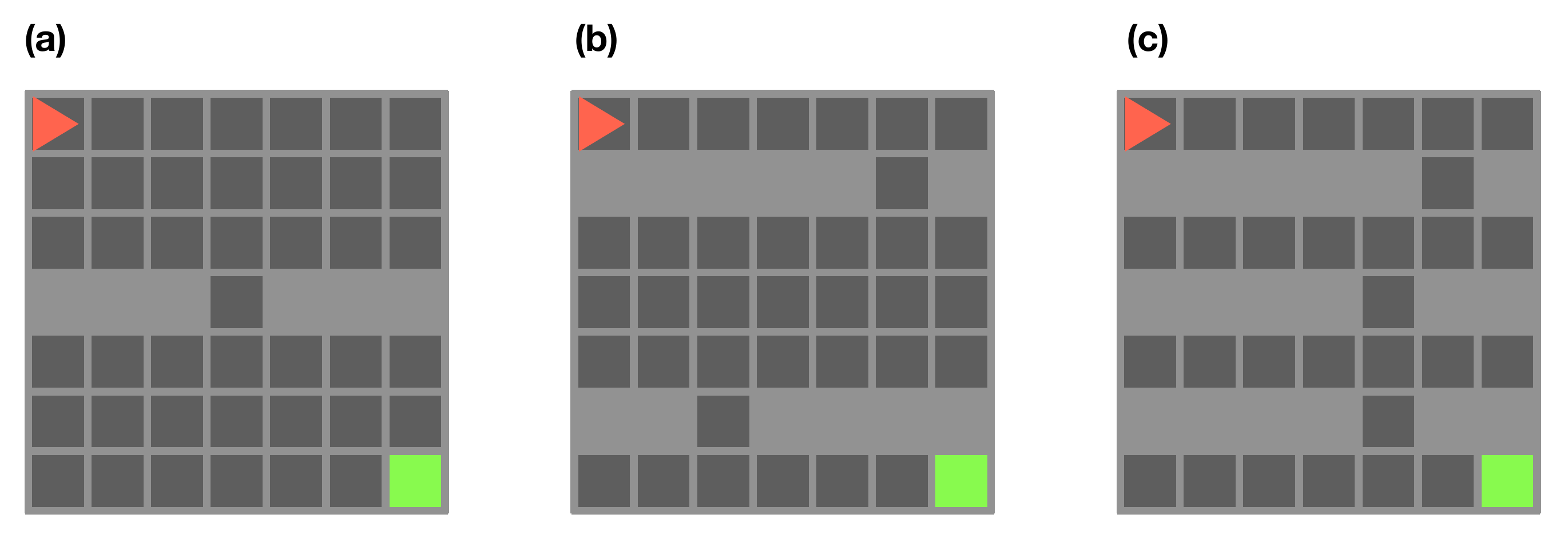}
\caption[Environment: MiniGrid-SimpleCrossing]{{\bfseries The SimpleCrossing environment from MiniGrid.} The three environments from MiniGrid-SimpleCrossing we consider in this work. In each environment, there are also walls which span $1$ unit on each side (not shown in the figure). (a), (b) and (c) represent examples from the MiniGrid-SimpleCrossingS9N1-v0, MiniGrid-SimpleCrossingS9N2-v0 and MiniGrid-SimpleCrossingS9N3-v0 environments, respectively.
 }
\label{SimpleCrossing_Env}
\end{figure}
\subsection{Hyperparameters and Model Size}
%
In the proposed QA3C, we use the Adam optimizer with learning rate $1 \times 10^{-4}$, $\beta_{1} = 0.92$ and $\beta_{2} = 0.999$. The local agents will update the parameters with the global shared memory every $S = 5$ steps. The discount factor $\gamma$ is set to be $0.9$. 
For the VQC, we set the number of qubits to be $8$ and two variational layers are used. Therefore, for each VQC, there are $8 \times 3 \times 2 = 48$ quantum parameters. Actor and critic both have their own VQC, thus the total number of quantum parameters is 96. The VQC architecture are the same across various testing environments considered in this work.
As we described in the \sectionautorefname{\ref{sec:QuantumA3C}}, single layer networks are used before and after the VQC to convert the dimensions of data. The networks preceding the VQC have input dimensions based on the environments that the agent is to solve. 
For the classical benchmarks, we consider the model which are very similar to the dressed VQC model. Specifically, we keep the architecture of classical model similar to the one presented in \figureautorefname{\ref{fig:hybrid_vqc}} while we replace the 8-qubit VQC with a single layer with input and output dimensions equal to 8. This makes the architecture very similar to the quantum model and the number of parameters are also very close. 
We summarize the number of parameters in \tableautorefname{\ref{tab:number_of_parameters}}.
\begin{table}[htbp]
\begin{tabular}{|l|ccc|c|}
\hline
\multirow{2}{*}{} & \multicolumn{3}{c|}{QA3C}                                             & Classical \\ \cline{2-5} 
                  & \multicolumn{1}{c|}{Classical} & \multicolumn{1}{c|}{Quantum} & Total & Total     \\ \hline
Acrobot           & \multicolumn{1}{c|}{148}       & \multicolumn{1}{c|}{96}      & 244   & 292       \\ \hline
Cart-Pole         & \multicolumn{1}{c|}{107}       & \multicolumn{1}{c|}{96}      & 203   & 251       \\ \hline
SimpleCrossing    & \multicolumn{1}{c|}{2431}      & \multicolumn{1}{c|}{96}      & 2527  & 2575      \\ \hline
\end{tabular}
\caption{{\bfseries Number of parameters.} 
We provide details on the number of parameters in the proposed QA3C model, which includes both quantum and classical components. The classical benchmarks were designed with architectures similar to the quantum model, resulting in similar model sizes.
}
\label{tab:number_of_parameters}
\end{table}
We utilize the open-source PennyLane package \cite{bergholm2018pennylane} to construct the quantum circuit models and the PyTorch as a overall machine learning framework. The number of CPU cores and hence the number of parallel agents is 80 in this work. We present simulation results in which the scores from the past 100 episodes are averaged.
\subsection{Results}
\subsubsection{Acrobot}
We begin by evaluating the performance of our models on the Acrobot environment. The simulation results of this experiment are presented in \figureautorefname{\ref{fig:results_acrobot}}. The total number of episodes was 100,000. As shown in the figure, the quantum model exhibits a gradual improvement during the early training episodes, while the classical model struggles to improve its policy. In terms of average score, the quantum model demonstrates superior performance compared to the classical model. Furthermore, the quantum model exhibits a more stable convergence pattern, without significant fluctuations or collapses after reaching optimal scores. These results suggest that the quantum model may be more robust and reliable in this environment.
\begin{figure}[htbp]
\includegraphics[width=0.5\linewidth]{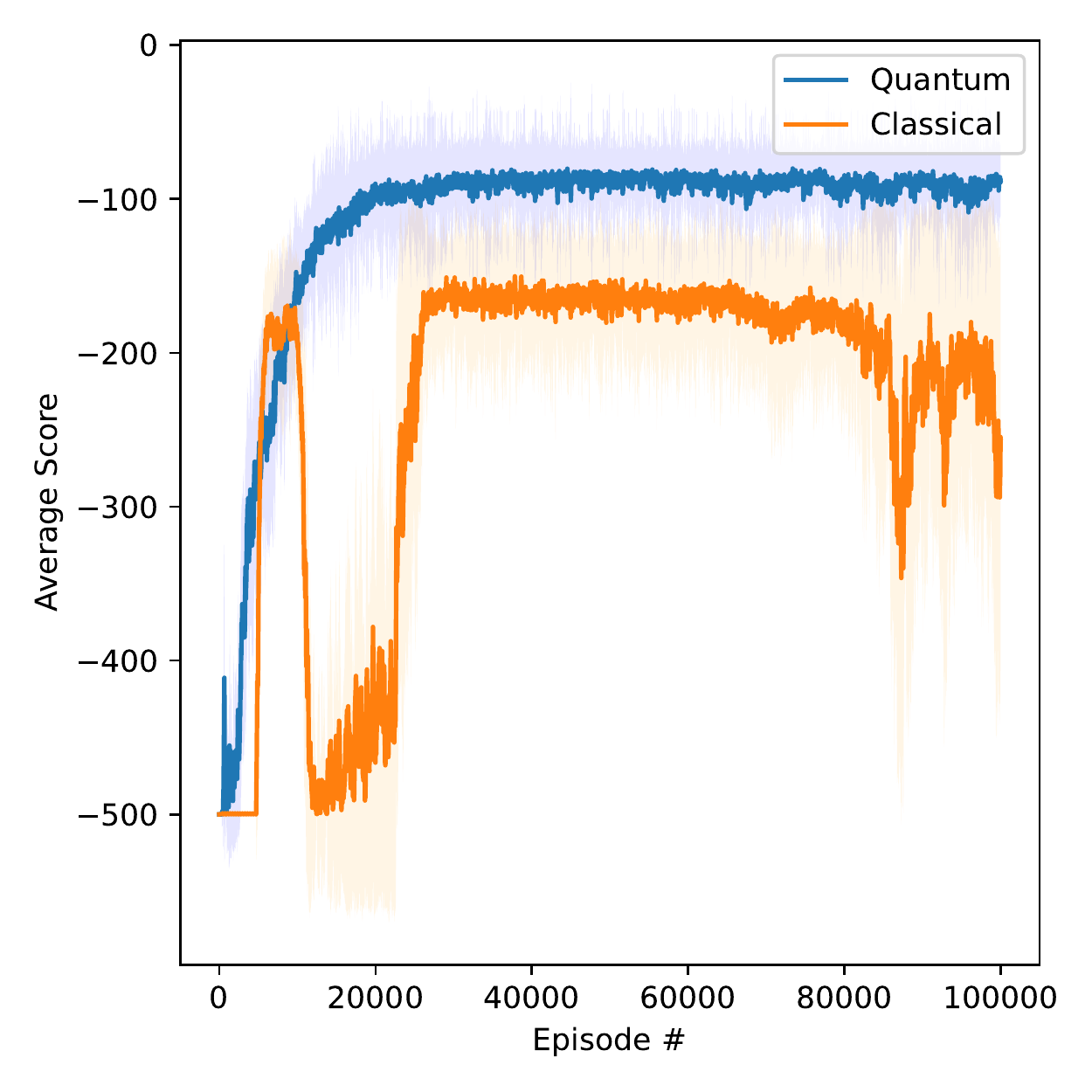}
\caption{{\bfseries Results: Quantum A3C in the Acrobot environment.}  }
\label{fig:results_acrobot}
\end{figure}
\subsubsection{Cart-Pole}
The next experiment was conducted in the Cart-Pole environment. The total number of episodes was 100,000. As illustrated in \figureautorefname{\ref{fig:results_cartpole}}, the quantum model achieved significantly higher scores than the classical model. While the classical model demonstrated faster learning in the early training episodes, the quantum model eventually surpassed it and reached superior scores. These results suggest that the quantum model may be more effective in this environment.
\begin{figure}[htbp]
\includegraphics[width=0.5\linewidth]{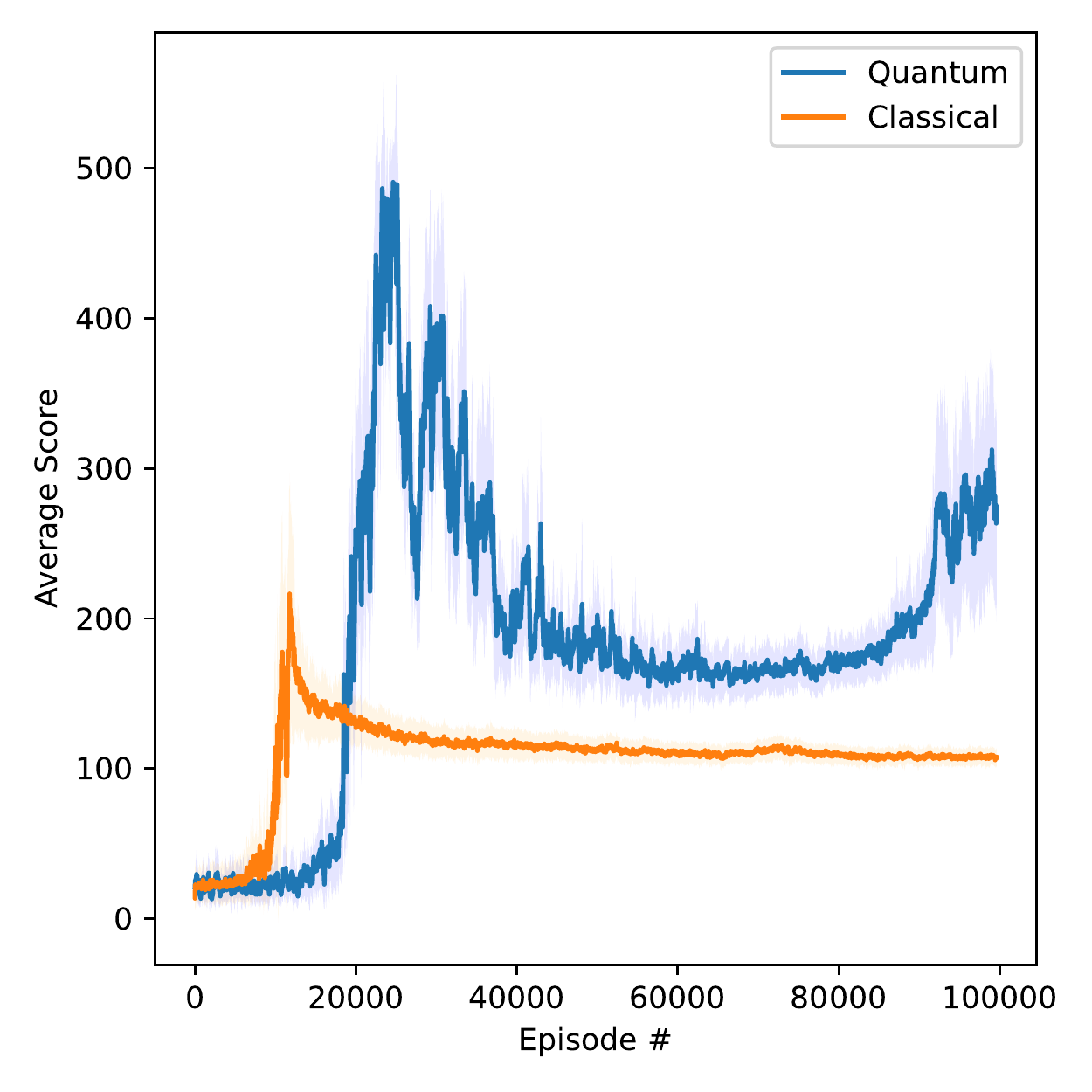}
\caption{{\bfseries Results: Quantum A3C in the CartPole environment.}  }
\label{fig:results_cartpole}
\end{figure}
\subsubsection{MiniGrid-SimpleCrossing}
The final experiment was conducted in the MiniGrid-SimpleCrossing environment, comprising a total of 100,000 episodes. As depicted in \figureautorefname{\ref{fig:results_simplecrossing}}, among the three scenarios, 
MiniGrid-SimpleCrossingS9N1-v0, MiniGrid-SimpleCrossingS9N2-v0, and MiniGrid-SimpleCrossingS9N3-v0,
the quantum model outperformed the classical model in two of the three scenarios, MiniGrid-SimpleCrossingS9N2-v0 and MiniGrid-SimpleCrossingS9N3-v0, demonstrating faster convergence and higher scores. Even in the remaining scenario, MiniGrid-SimpleCrossingS9N1-v0, the difference in performance between the two models was minor.
\begin{figure}[htbp]
\includegraphics[width=1\linewidth]{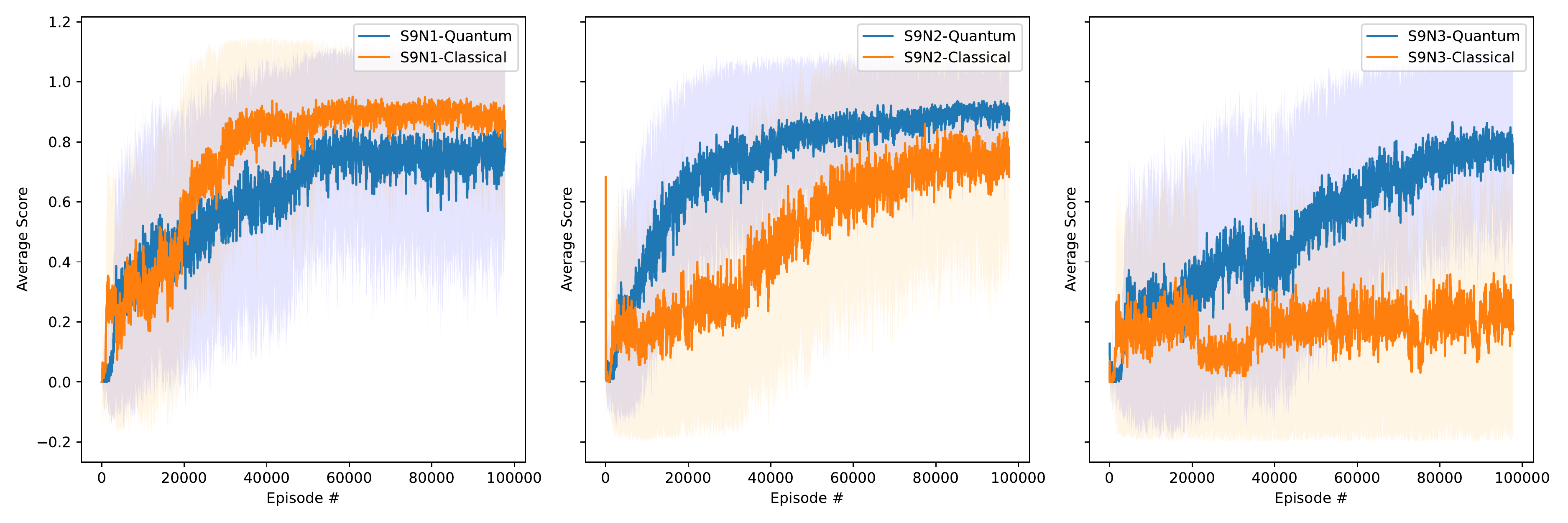}
\caption{{\bfseries Results: Quantum A3C in the MiniGrid-SimpleCrossing environment.}  }
\label{fig:results_simplecrossing}
\end{figure}
%

\section{\label{sec:Conclusion}Conclusion}
In this study, we demonstrate the effectiveness of an asynchronous training framework for quantum RL agents. Through numerical simulations, we show that in the benchmark tasks considered, advantage actor-critic quantum policies trained asynchronously can outperform or match the performance of classical models with similar architecture and sizes. This technique affords a strategy for expediting the training of quantum RL agents through parallelization, and may have potential applications in various real-world scenarios.

\begin{acknowledgments}
The views expressed in this article are those of the authors and do not represent the views of Wells Fargo. This article is for informational purposes only. Nothing contained in this article should be construed as investment advice. Wells Fargo makes no express or implied warranties and expressly disclaims all legal, tax, and accounting implications related to this article.
\end{acknowledgments}

\appendix

\section{Algorithms}
\subsection{Quantum-A3C}
\begin{center}
\scalebox{0.9}{
\begin{minipage}{\linewidth}

\begin{algorithm}[H]
\begin{algorithmic}
\State \textbf{Define} the global update parameter $S$
\State \textbf{Assume} global shared hybrid VQC policy parameter $\theta$
\State \textbf{Assume} global shared hybrid VQC value parameter $\theta_{v}$
\State \textbf{Assume} global shared episode counter $T = 0$
\State \textbf{Assume} process-specific hybrid VQC policy parameter $\theta'$
\State \textbf{Assume} process-specific hybrid VQC value parameter $\theta_{v}'$
\State \textbf{Initialize} process-specific counter $t = 1$

\While{$T < T_{max}$}
    \State Reset gradients $d \theta \leftarrow 0$ and $d \theta_{v} \leftarrow 0$
    \State Set $t_{start} = t$
    \State Reset the environment and get state $s_{t}$
    \While{$s_{t}$ non-terminal or $t - t_{start} < t_{max}$}
        \State Perform $a_{t}$ according to policy $\pi(a_{t}|s_{t};\theta')$
        \State Receive reward $r_{t}$ and the new state $s_{t+1}$
        \State Update process-specific counter $t \leftarrow t+1$
        \If{$t \mod S = 0$ or reach terminal state}
            \State Set $R=\left\{\begin{array}{l}
            0 \quad \text{for terminal $s_{t}$}\\ 
            V\left(s_t, \theta_v^{\prime}\right) \quad \text{for non-terminal $s_{t}$}
            \end{array}\right.$
            \For{$i \in\left\{t-1, \ldots, t_{\text {start }}\right\}$}
                \State $R \leftarrow r_i+\gamma R$
                \State Accumulate gradients wrt $\theta'$: $d \theta \leftarrow d \theta+\nabla_{\theta^{\prime}} \log \pi\left(a_i \mid s_i ; \theta^{\prime}\right)\left(R-V\left(s_i ; \theta_v^{\prime}\right)\right)$
                \State Accumulate gradients wrt $\theta_{v}'$: $d \theta_v \leftarrow d \theta_v+\partial\left(R-V\left(s_i ; \theta_v^{\prime}\right)\right)^2 / \partial \theta_v^{\prime}$
            \EndFor
            \State Perform asynchronous update of $\theta$ using $d\theta$ and of $\theta_{v}$ using $d\theta_{v}$
            \State Update process-specific parameters from global parameters: $\theta' \leftarrow \theta$ and $\theta_{v}' \leftarrow \theta_{v} $
        \EndIf
    \EndWhile
\EndWhile

\end{algorithmic}
\caption{Quantum asynchronous advantage actor-critic learning (algorithm for each actor-learner process)}
\label{QA3C_alg}
\end{algorithm}
\end{minipage}
}
\end{center}
\bibliographystyle{ieeetr}
\bibliography{apssamp,bib/classical_rl,bib/vqc,bib/qc_basic,bib/tool,bib/qml_examples,bib/qrl}

\end{document}